\documentclass[journal,twoside,web]{ieeecolor}
\usepackage{generic}
\usepackage{cite}
\usepackage{amsmath,amssymb,amsfonts}
\usepackage{algorithmic}
\usepackage{graphicx}
\usepackage{algorithm,algorithmic}
\usepackage{hyperref}
\hypersetup{hidelinks=true}
\usepackage{textcomp}
\usepackage{stfloats}
\usepackage{url}
\usepackage{verbatim}
\usepackage{graphicx}
\usepackage{multirow}
\usepackage{booktabs}
\usepackage{colortbl}
\def\BibTeX{{\rm B\kern-.05em{\sc i\kern-.025em b}\kern-.08em
    T\kern-.1667em\lower.7ex\hbox{E}\kern-.125emX}}
\begin{document}

\title{Unpaired Optical Coherence Tomography Angiography Image Super-Resolution via Frequency-Aware Inverse-Consistency GAN}

\author{Weiwen Zhang, Dawei Yang, Haoxuan Che, \IEEEmembership{Student Member, IEEE}, An Ran Ran, Carol Y. Cheung, and Hao Chen, \IEEEmembership{Senior Member, IEEE}
\thanks{This work was supported by funding from HKUST 30 for 30 Research Initiative Scheme, Hong Kong University of Science and Technology, and Shenzhen Science and Technology Innovation Committee (Project No. SGDX20210823103201011), and Direct Grants from The Chinese University of Hong Kong (Project Code: 4054419 \& 4054487).  \textit{(Corresponding author: Hao Chen.)}}
\thanks{Weiwen Zhang and Haoxuan Che are with the Department of Computer Science and Engineering, The Hong Kong University of Science and Technology, Hong Kong, China (e-mail: wzhangbu@cse.ust.hk; hche@cse.ust.hk).}
\thanks{Dawei Yang, An Ran Ran, and Carol Y. Cheung are with the Department of Ophthalmology and Visual Sciences, The Chinese University of Hong Kong, Hong Kong, China (e-mail: gabrielyang@link.cuhk.edu.hk; emma\_anran@link.cuhk.edu.hk; carolcheung@cuhk.edu.hk).}
\thanks{Hao Chen is with the Department of Computer Science and Engineering, The Hong Kong University of Science and Technology, Hong Kong, China, and also with the Department of Chemical and Biological Engineering, The Hong Kong University of Science and Technology, Hong Kong, China (e-mail: jhc@cse.ust.hk). }
}

\maketitle

\begin{abstract}
  For optical coherence tomography angiography (OCTA) images, the limited scanning rate leads to a trade-off between field-of-view (FOV) and imaging resolution. Although larger FOV images may reveal more parafoveal vascular lesions, their application is hampered due to lower resolution. To increase the resolution, previous works only achieved satisfactory performance by using paired data for training, but real-world applications are limited by the challenge of collecting large-scale paired images. Thus, an unpaired approach is highly demanded. Generative Adversarial Network (GAN) has been commonly used in the unpaired setting, but it may struggle to accurately preserve fine-grained capillary details, which are critical biomarkers for OCTA. In this paper, our approach aspires to preserve these details by leveraging the frequency information, which represents details as high-frequencies ($\textbf{\textit{hf}}$) and coarse-grained features as low-frequencies ($\textbf{\textit{lf}}$). We propose a GAN-based unpaired super-resolution method for OCTA images and exceptionally emphasize $\textbf{\textit{hf}}$ fine capillaries through a dual-path generator. To facilitate a precise spectrum of the reconstructed image, we also propose a frequency-aware adversarial loss for the discriminator and introduce a frequency-aware focal consistency loss for end-to-end optimization. We collected a paired dataset for evaluation and showed that our method outperforms other state-of-the-art unpaired methods both quantitatively and visually. Code can be accessed at: https://github.com/KevynUtopia/FUSR.
\end{abstract}


 \begin{figure}[h!] 
    \centering 
    \includegraphics[scale=.12]{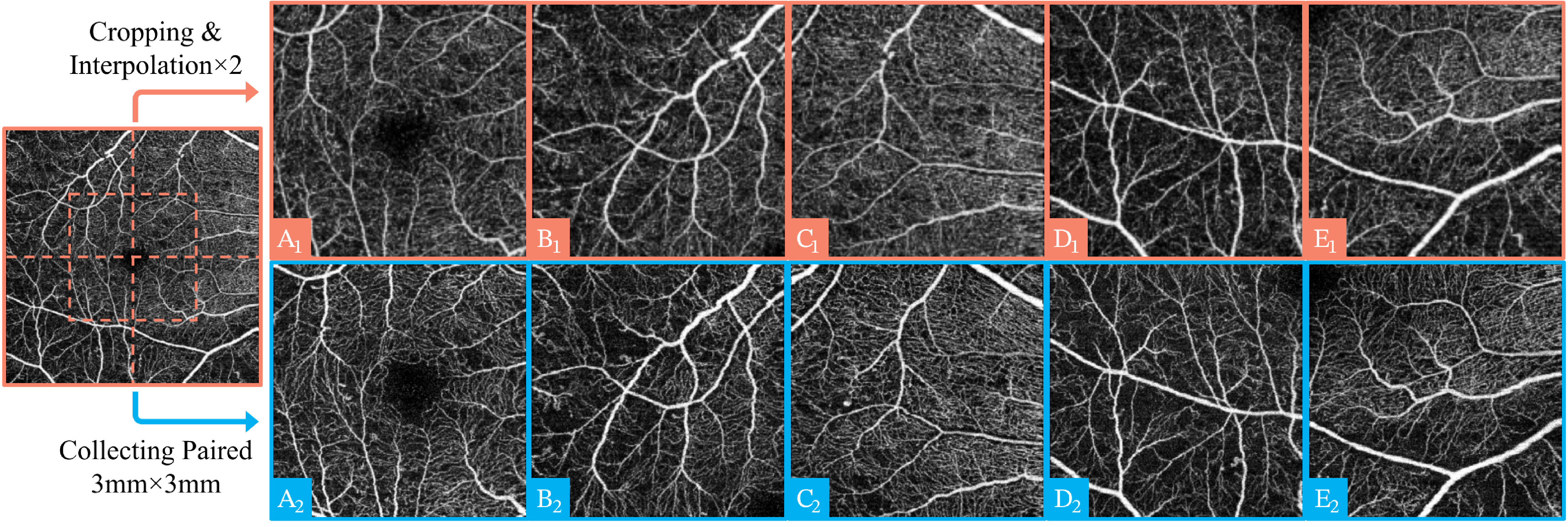} 
    \caption{Illustration for our OCTA images dataset, which is retrospectively collected from the Chinese University of Hong Kong Sight-THReatening Diabetic Retinopathy (CUHK-STDR) study. Leftmost illustrates the original 6mm$\times$6mm image. Orange boxes with subscript 1 indicate 6mm$\times$6mm patches. Blue boxes with subscript 2 indicate the corresponding paired 3mm$\times$3mm images. The same letters indicate the same area. A: Fovea-center images. B$\sim$E: Parafoveal images.}
    \label{data} 
\end{figure}

\section{Introduction}

\IEEEPARstart{O}{ptical} coherence tomography angiography (OCTA) is an imaging modality based on the optical coherence tomography (OCT) platform, which generates depth-resolved images of the retina and choroidal microvasculature~\cite{jia2015quantitative}. OCTA can support the evaluation of multiple retinal diseases, including diabetic retinopathy and age-related macular degeneration~\cite{hwang2015optical,
roisman2016optical,sun2019oct,che2022learning,che2023dgdr,yang2023assessment}. However, due to the limited scanning rate of commercial OCT instruments~\cite{spaide2018optical,article1}, images with a smaller field-of-view (FOV) have higher axial scan density and thus higher resolution~\cite{gao2016optical}. As a result, among most common FOV, 3mm$\times$3mm (Fig. \ref{data}. $A_{1}$$\sim$$E_{1}$) is more widely employed in clinical settings to visualize finer capillaries, as compared to the 6mm$\times$6mm~\cite{Yangbjophthalmol-2021-320779,yang2023assessment,hormel2021artificial,waheed2023optical}. Nevertheless, larger FOV (Fig. \ref{data}. $A_{2}$) is supposed to reveal more parafoveal vascular lesions~\cite{nafar2021clinical}. Therefore, improving the resolution of 6mm$\times$6mm images (Fig. \ref{data}. $A_{2}$), on par with 3mm$\times$3mm images, will further empower ophthalmologists to evaluate capillary losses and develop more personalized treatments~\cite{ho_2019_comparison,ziqi2022using,che2023iqad}. In computer vision, this task refers to super-resolution which upscales images (Fig. \ref{data}. $A_{2}$ to $B_{2} \sim E_{2}$) and improves the quality through restoration.

For OCTA image super-resolution, most previous works adopt the paired setting for training to achieve the desired qualitative performance~\cite{article1,article2,hao2022sparse}. One approach involves collecting and creating paired high-resolution ($HR$) and low-resolution ($LR$) images from the same eye of the same patient~\cite{article1,article2}. However, collecting large-scale real-paired images requires sophisticated image registration, which is laborious and challenging and may hinder the medical application~\cite{paavilainen2021bridging}. Another approach creates pseudo image pairs using bicubic interpolation to synthesize $LR$ from $HR$ for training~\cite{hao2022sparse}. However, interpolation is an oversimplified presumption since it may not accurately represent real-world degradation. Alternatively, the unpaired OCTA image super-resolution could potentially serve as a mitigation. Therefore, we propose an unpaired approach by jointly optimizing the restoration with a degradation model~\cite{bulat2018learn}. Also regarding the consensual merits of Generative Adversarial Networks (GANs), the models are accordingly formulated and optimized via consistency loss~\cite{goodfellow2020generative,zhao2018unsupervised,yuan2018unsupervised,maeda2020unpaired,zhu2017unpaired}. 

\begin{figure}[t]
    \centering
    \includegraphics[scale=0.123]{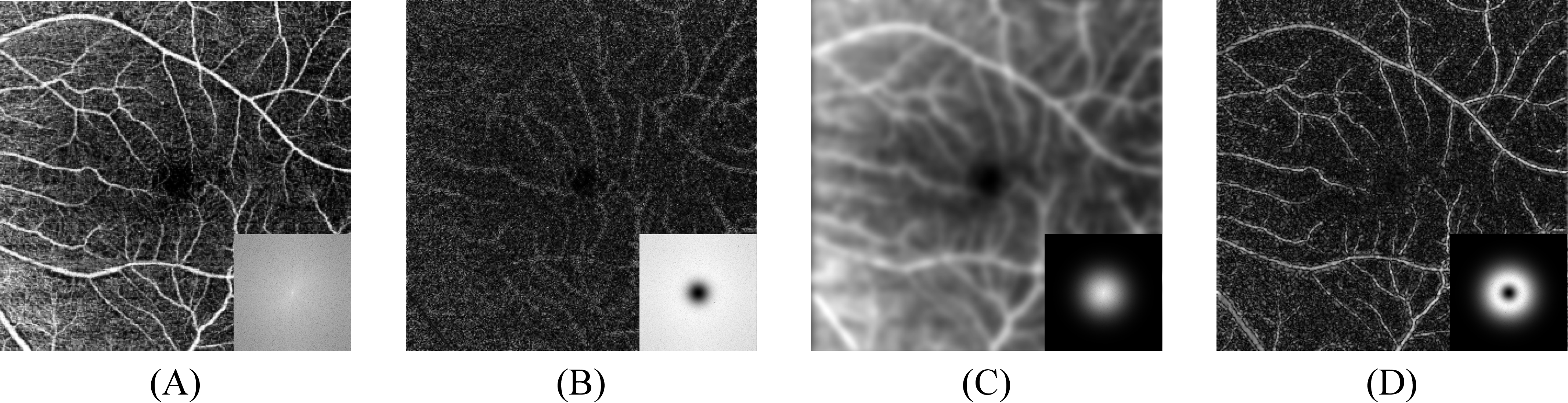}
    \caption{Illustration for 6mm$\times$6mm OCTA images and frequency components. 
    Lower-right corners are bandwidth spectral filters. A: 6mm$\times$6mm OCTA image and its spectrum. 
    B: \textbf{$hf$} and its high-pass filter. 
    C: \textbf{$lf$} and its low-pass filter. 
    D: Middle-frequencies and its middle-pass filter.}
\label{freq_img}
\end{figure}

Moreover, higher resolution of the capillary network will allow more accurate assessments of eye diseases related to microvasculature~\cite{xie2023retinal,9284503,qian2024drac}. Thus the algorithm should exceptionally emphasize the fine-grained vessels. In the frequency domain, these capillaries correspond to high-frequency ($hf$) information (Fig. \ref{freq_img}. B), whereas the general illuminance and colors correspond to low frequencies ($lf$) (Fig. \ref{freq_img}. C). However, convolutional neural networks (CNNs) inherently exhibit a bias towards $lf$ \cite{xu2019frequency}. This bias can also be observed in the spectral distribution (Fig. \ref{discrepancy}), which illustrates a discernibly increasing discrepancy between reconstructed images and $HR$ ground truths as the bandwidth increases. Though super-resolution aspires to enhance $hf$ details, such bias may result in inaccurate or deficient details~\cite{wang2021unsupervisedDA,ji2021frequency,tai2017image}. Consequently, for OCTA images, capillary structures in microvasculature might be altered. To alleviate these issues, our approach directly leverages frequency information and imposes exceptional emphasis on $hf$, aiming at accurate and sufficient fine-grained details. Specifically in our restoration and degradation GAN, to preserve salient $hf$ details in generators, we separate frequency components in a dual-path structure for feature extraction and then fuse features for reconstruction via residual blocks~\cite{he2016deep}. To also facilitate discriminators being sensitive to $hf$, we introduce the frequency-aware adversarial loss (\textbf{FAL}) to consider both frequency and spatial components. To consistently preserve the frequency distribution, we propose a frequency-aware focal consistency loss (\textbf{FFCL}).

In general, by leveraging frequency information and imposing exceptional emphasis on $hf$, this paper introduces the Frequency-aware Unpaired Super-Resolution for OCTA images (\textbf{FUSR}) by extending our previous work published in Medical Image Computing and Computer Assisted Intervention (MICCAI)~\cite{zhang2022frequency}. To surmount $lf$-bias of neural networks, we propose a dual-path architecture in generators to separately refine $hf$ capillary details. To also facilitate discriminators being aware of frequencies, we propose \textbf{FAL} by exploiting wavelet space. To further preserve spectral distribution, we propose the \textbf{FFCL} to penalize spectral errors. Then, our approach exploits both frequency and spatial domains to effectively produce high-resolution images in the unpaired setting.

This paper contributes the following perspectives: 
\begin{itemize} 
     \item To resolve the unpaired OCTA super-resolution, we propose a GAN-based approach containing restoration and degradation models, and jointly optimize it using consistency losses in an end-to-end manner.

     \item To mitigate $lf$-bias and to enhance fine-grained details, we exceptionally emphasize $hf$ components through a dual-path structure. We also propose the frequency-aware adversarial loss for restoration and degradation GANs, and the frequency-aware focal consistency loss to preserve the spectral consistency.

     \item To quantitatively evaluate the performance, we purposely collect fovea-central and parafoveal paired $HR$ and $LR$ images from CUHK-STDR study for different paired metrics, and verify the superiority of our method.
\end{itemize}

\section{Related works}
\subsection{Image Super-Resolution}
Super-Resolution is a prominent task in low-level computer vision, aimed at increasing the resolution and restoring imaging quality, including OCTA images~\cite{article1}. In recent years, with the advancements in deep learning, learning-based techniques have demonstrated remarkable performance, particularly through supervised learning using paired datasets \cite{dong2014learning}. However, for most imaging modalities in the real world, the challenge of collecting large-scale paired datasets remained a fundamental issue \cite{wei2021unsupervised}. To overcome this challenge, a common approach was to generate training data by downsampling $HR$ images to synthetic $LR$ counterparts using interpolation \cite{dong2014learning}. Subsequently, algorithms aimed to recover the super-resolution models that restore $LR$. However, interpolation-based downsampling oversimplified the degradation process, leading to models that may not generalize well to real-world super-resolution. To relieve the issues of the paired setting, GAN has been introduced~\cite{ledig2017photo}. Adversarial loss guided the model to produce more visually pleasing results~\cite{goodfellow2020generative}, but these methods still relied on paired data for training.

Alternatively, unpaired methods explored the restoration by formulating the degradation using unpaired images. These methods could be categorized into two main approaches: \textit{\textbf{two-stage}} and \textit{\textbf{one-stage}} methods. \textit{\textbf{Two-Stage}} methods primarily formulated the degradation process, followed by a separate optimization of the restoration model. \cite{ji2020real} proposed a kernel estimation approach to mimic real-world degradation. \cite{bulat2018learn} introduced a High-to-Low GAN, which generates the $LR$ from $HR$. To ensure stable optimization of the degradation, \cite{chen2020unsupervised} proposed to synthesize $LR$ and utilized unsupervised learning to bridge the gap between real and synthesized images. However, for two-stage approaches, the restoration might be highly affected by the performance of degradation, causing suboptimal restoration results~\cite{wang2020deep}.

\textit{\textbf{One-Stage}} approaches aimed to jointly optimize the models in an end-to-end manner~\cite{zhu2017unpaired}, which commonly employed GAN and consistency loss. For instance, a bi-cycle network was proposed to jointly generate real-world $LR$ and optimize the super-resolution model~\cite{zhao2018unsupervised}. However, GAN has suffered from instability in the training phase and thus is prone to introduce unexpected noises~\cite{yamaguchi2021f,khayatkhoei2022spatial}. To this end, several solutions were proposed. Some modified the architecture of the pipeline. \cite{yuan2018unsupervised} proposed a cycle-in-cycle structure using nested GANs and consistency losses. \cite{maeda2020unpaired} proposed a pseudo-supervision using corrected-clean and pseudo-clean $LR$ as intermediates between $LR$ and $HR$ images. Another attempt was to introduce frequency information. By exploiting both spatial and wavelet domains, \cite{wei2021unsupervised} resolved unpaired super-resolution via domain adaptation to tackle the gap between real and synthetic images.

\begin{figure}[!t]
\centering
\includegraphics[scale=.235]{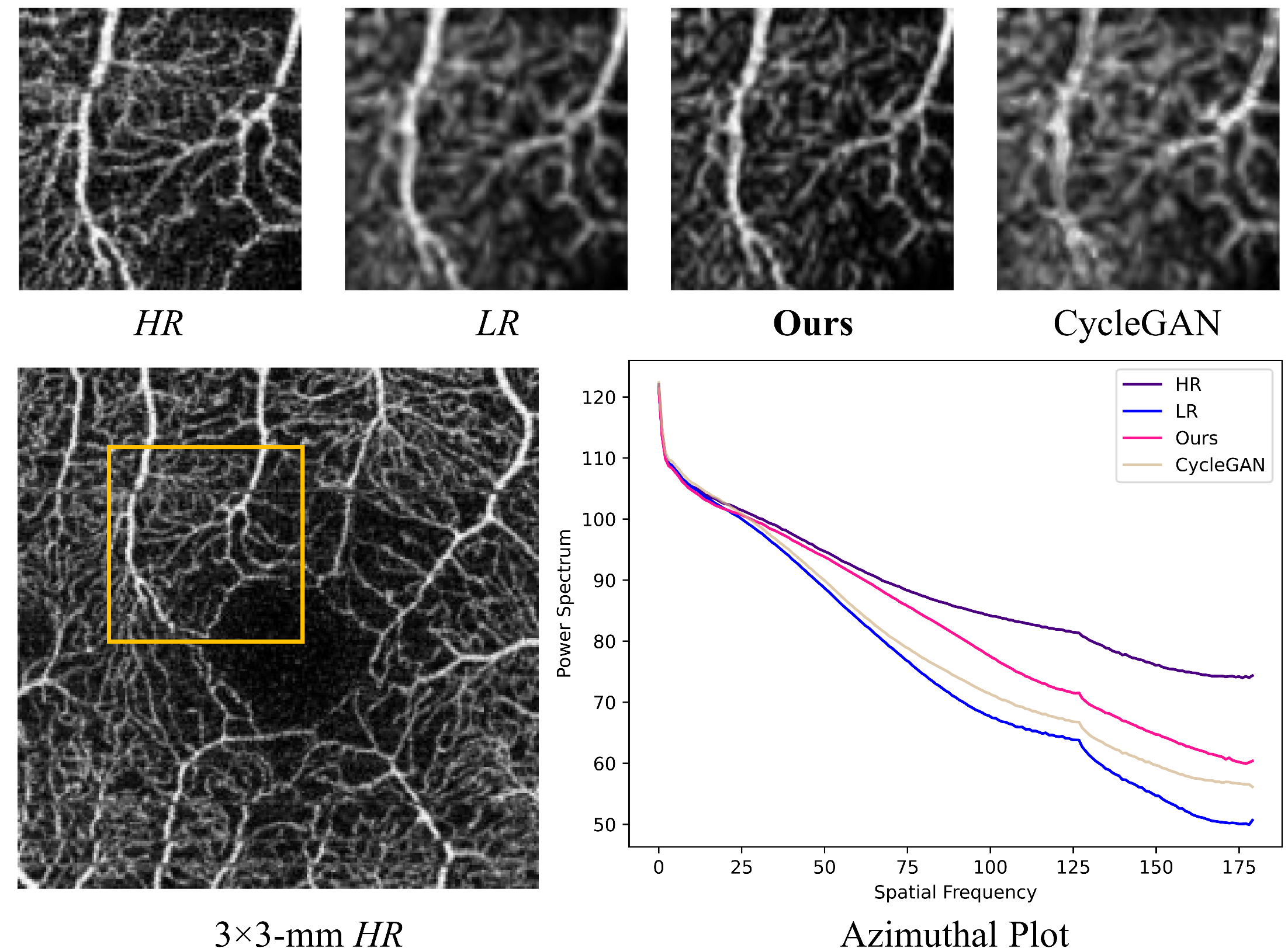}
\caption{Azimuthal integral on spectrum as specified in Eq. \eqref{azimuthal}. It indicates that $HR$ contains stronger power in middle- and high-bands of the spectrum than $LR$. While the middle frequencies of different methods are similar, our approach better fits the $lf$ information and enhances $hf$ information compared to the real $HR$.}
\label{discrepancy}
\end{figure}

\subsection{Super-Resolution for OCTA Images}
OCTA Super-Resolution algorithms specifically aimed to upscale 6mm$\times$6mm images and improve their quality on a par with 3mm$\times$3mm images. However, the preparation of training OCTA images posed a laborious challenge \cite{paavilainen2021bridging}. Existing methods addressed this challenge by either synthesizing $LR$ or collecting paired images. For example, \cite{hao2022sparse} proposed a method that degrades $HR$ images through interpolation downsampling. Although this approach intended to mitigate domain gaps between real- and generated images using GAN, the restoration modeling overlooked complex real-world degradation, which could be influenced by different FOV, limited scanning rates, etc. Alternative methods veritably collected paired $HR$ and $LR$ images from the same eye of the same patient and employed a supervised approach to enhance the $LR$ images~\cite{article1,article2}. However, these approaches suffered from two inherent limitations when applied to OCTA images. Firstly, to prepare pixel-wise paired images for training and evaluation, registration should be used to mitigate structural changes from image capturing. Such preprocessing inevitably altered the original structure within the OCTA image, resulting in unconvincing supervision. Secondly, due to different FOV, each 3mm$\times$3mm image could only provide incomplete supervision with only a sub-region of $HR$ information for each 6mm$\times$6mm image. The above limitations motivated our unpaired OCTA super-resolution, which could release the reliance on paired data and implicitly formulate the restoration and degradation using GAN.

\subsection{Frequency-Domain Analysis}
It had been shown that deep neural networks tend to fit $lf$ more precisely than $hf$ \cite{xu2019frequency}. Especially for GANs that are commonly utilized in unpaired super-resolution, models also suffered from the bias and resulted in missing $hf$ details or unexpected artifacts~\cite{yamaguchi2021f,khayatkhoei2022spatial}. The spectral distribution can also demonstrate this bias, where $hf$ were not reconstructed as sufficiently as $lf$ (see Fig. \ref{discrepancy}). To produce high-fidelity $HR$ images, $hf$ capillary details \cite{cai2021frequency} should have been meticulously preserved. To address this inherent gap, frequency information has been included in deep learning frameworks by separating frequency components \cite{xu2020Learning}. \cite{fritsche2019frequency} leveraged frequency components in unpaired super-resolution algorithm. \cite{liu2023spectral} formulated dual-domain learning to estimate the spectral uncertainty and to enhance the perceptual quality in the super-resolution. In addition to incorporating frequency in the frameworks, several frequency-aware losses have been introduced to yield more realistic results. \cite{wang2021unsupervisedDA,li2022wavelet} proposed exploiting the wavelet domain to mitigate the domain gap between real and synthetic images. \cite{ji2021frequency} introduced Frequency Consistent Adaptation to ensure frequency domain consistency. The focal frequency loss \cite{jiang2021focal} adaptively focused on frequency components using amplitude and phase information. \cite{cai2021frequency} suggested a Fourier frequency loss to separately preserve high and low-frequency amplitudes. However, studies that use frequency and spatial information for unpaired OCTA super-resolution have not been sufficiently conducted yet.

To precisely restore OCTA $LR$ images using unpaired data, this paper proposes leveraging spatial and frequency components in the restoration and degradation frameworks to benefit unpaired super-resolution. To reconstruct precise frequency information, we introduce Frequency-aware Adversarial Loss for discriminators and the Frequency-aware Focal Consistency Loss for our end-to-end framework.

\begin{figure*}[t]
\centering
\includegraphics[scale=0.191]{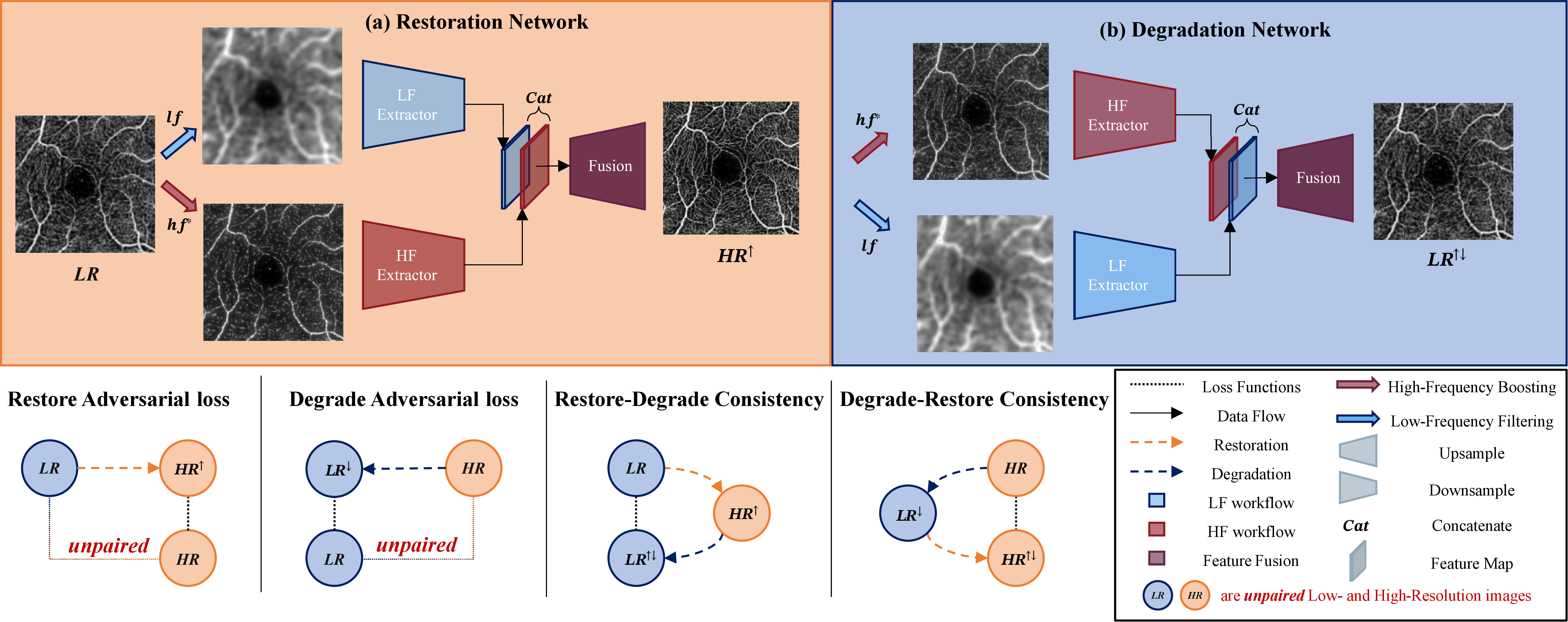}
\caption{An overview of our methods. 
The input $LR$ image is decomposed into $lf$ and $hf^*$ (through $HFB$) and fused for restoring to $HR^{\uparrow}$. Then, it is degraded to $LR^{\uparrow\downarrow}$ and optimized through restoration-degradation consistency and adversarial losses. The restoration network is eventually taken as the OCTA high-resolution model in the inference phase. The inverse degradation-restoration process is represented in simplified conceptual graphs. Note that input $LR$ and $HR$ images are unpaired in the training phase.
}
\label{fig:pipeline}
\end{figure*}

\begin{table}[b]
    \centering

\caption{The summary of the abbreviations used in this paper.}
    \begin{tabular}{l>{\centering\arraybackslash}p{1.2cm}>{\centering\arraybackslash}p{5.8cm}}
    \toprule
    & Abbreviation & Meaning \\
    \hline
    \multirow{3}{*}{\rotatebox{90}{Method}} 
        &FUSR & \textbf{F}requency-aware \textbf{U}npaired \textbf{S}uper-\textbf{R}esolution\\
        &FFCL & \textbf{F}requency-aware \textbf{F}ocal \textbf{C}onsistency \textbf{L}oss \\
        &FAL & \textbf{F}requency-aware \textbf{A}dversarial \textbf{L}oss \\
    \hline
    \multirow{4}{*}{\rotatebox{90}{Notation}} 
        &FFT & \textbf{F}ast \textbf{F}ourier \textbf{T}ransformation \\
        &DWT & \textbf{D}iscrete \textbf{W}avelets \textbf{T}ransformation \\
        &{hf}, {lf} & \textbf{h}igh-, \textbf{l}ow-\textbf{f}requency \\
        &{HR}, {LR} & \textbf{H}igh-, \textbf{L}ow-\textbf{R}esolution \\
    \bottomrule
    \end{tabular}
    \label{table:abbr}

\end{table}
\section{Methodology}
To resolve unpaired super-resolution for OCTA images, we propose GAN-based restoration and inverse degradation models, which are optimized in an end-to-end manner through consistency loss~\cite{zhu2017unpaired}, as depicted in Fig. \ref{fig:pipeline}. To mitigate the frequency bias and thus precisely enhance $hf$ capillary details, we leverage different frequency components via a dual-path structure within the framework. In the GAN paradigm, we also exploit the frequency domain through \textbf{FAL} for the discriminators, and propose an \textbf{FFCL} to guarantee accurate spectrum as an objective for end-to-end learning. For convenient comprehension, we highlight the abbreviations with corresponding meanings in this paper in Table.~\ref{table:abbr}.

\subsection{Preliminaries}
3mm$\times$3mm OCTA images are defined as $HR$ and 6mm$\times$6mm images as $LR$. We represent the restoration process with the mapping from $LR$ to $HR$, denoted by \(G_{Res}: LR \rightarrow HR^{\uparrow}\), and the degradation process as the inverse mapping from $HR$ to $LR$, represented by \(G_{Deg}: HR \rightarrow LR^{\downarrow}\). Here, $HR^{\uparrow}$ and $LR^{\downarrow}$ refer to the generated $HR$ and $LR$ images.

Given an image $x\in \mathbb{R}^{\mathrm{M}\times\mathrm{N}}$ (can be either $HR$ or $LR$), its representation in the frequency domain is denoted as \(\mathcal{X}\) via Fast Fourier Transformation (FFT):
\begin{flalign}
    \label{fft}
    \mathcal{X}\left ( u,v \right )  &= \sum_{m=0}^{M-1}\sum_{n=0}^{N-1}x\left(m,n\right)e^{-i2\pi\left ( \frac{um}{M}+\frac{vn}{N}\right ) } \nonumber \\
    &= R\left( u,v \right) +iI\left( u,v \right)  
\end{flalign}
where $M$ and $N$ are image height and width, while $(m,n)$ and $(u,v)$ are Cartesian coordinates of the spatial- and frequency-domain pixels. The complex numbers in $\mathcal{X}$ produced by FFT can be further separated into real $(R)$ and imaginary parts $(I)$ with respect to the Euler's formula: $e^{i\theta}=cos\theta +isin\theta$.

Then, to visualize the spectral power of the image, the azimuthal integral over the spectrum~\cite{durall2020watch} is defined as:
\begin{flalign}
\label{azimuthal}
& A(\omega_k)=\int_{0}^{2\pi} \left \| \mathcal{X}(\omega_k\cdot \cos(\phi ), \omega_k\cdot \sin(\phi)) \right \|^{2}d\phi
\end{flalign}
where ($\phi$, $\omega_k$) is the azimuth and $k$-bandwidth in polar coordinate of the image, and $k=0,1,...,M/2-1$.

To leverage different frequency components, we define the kernel of Gaussian filters as:
\begin{flalign}
G\left( u,v \right) =\frac{1}{2\pi \sigma ^2}e^{-\left( u^2+v^2 \right) /\left( 2\sigma ^2 \right)}
\label{gaussianblur} 
\end{flalign}
where $\sigma$ is the variance of Gaussian kernel, and ($u$,$v$) is Cartesian coordinate. According to the convolution theorem (see Fig. \ref{freq_img}), Gaussian blurry in the spatial domain can be represented as low-pass filtering in the frequency domain. Thus the $lf$ is extracted by Gaussian blurring using convolution operation, $lf=G \ast x$. Then, $hf$ is obtained by subtracting the $lf$ information from the original image, \(hf = x-lf\). 

\subsection{Frequency-aware Restoration and Degradation}
Our unpaired super-resolution framework contains a restoration model $G_{Res}$ and a degradation model $G_{Deg}$. Models can be optimized in one-stage end-to-end manner according to adversarial loss and consistency loss~\cite{goodfellow2020generative,zhu2017unpaired}.

In spatial domain, super-resolution aims to effectively and accurately refine fine-grained details. In the frequency domain, these details are widely recognized as $hf$, while general contours are referred to as $lf$~\cite{li2022wavelet}. But limited by the inherent bias of neural networks, $hf$ may not be sufficiently boosted~\cite{xu2019frequency}. Therefore, our work proposes a frequency-aware architecture that leverages both spatial and frequency information, as shown in Fig. \ref{fig:pipeline}. 

Specifically, to emphasize $hf$ and alleviate the bias to $lf$, we intentionally separate frequency components using Gaussian filtering, as shown in Eq. \eqref{gaussianblur}. The corresponding feature maps are embedded through a dual-path structure (see Fig. \ref{fig:pipeline}(a) and (b)). Then, the image is reconstructed by concatenating and fusing these features through a trainable fusion module. In such manner, $G_{Res}$ enhances $hf$ details while preserving $lf$ coarse-grained features, and $G_{Deg}$ filters out $hf$ while retaining $lf$ components. Additionally, to provide $hf$ for 1 without breaking vessel coherence, we define an operation dubbed as the high-frequency boosting ($HFB$), in the following form:
\begin{flalign}
hf^{*}= x + \alpha*hf 
\label{hfb} 
\end{flalign}
where $hf^{*}$ represents the boosted $hf$, and $\alpha$ is the factor determining the extent of enhancement. Since providing pure $hf$ to the network may break vessel structures and cause incoherence, we utilize the $HFB$ to obtain $hf^{*}$, shown in Eq. \eqref{hfb}. Then $lf$ and $hf^{*}$ are provided as frequency components to the dual-path generators for feature extraction via residual blocks~\cite{he2016deep}. Subsequently, the features from the two paths are fused for the image reconstruction.

To optimize the $G_{Res}$ using the unpaired dataset, we incorporate consistency loss to preserve the vessel structures, namely inverse consistency since restoration and degradation represent inverse mappings between $LR$ and $HR$. We formulate the degradation-restoration inverse-consistency, \(G_{Deg}\cdot G_{Res}: LR \rightarrow HR^{\uparrow} \rightarrow LR^{\uparrow\downarrow}\), using the \(L_{1}\) norm loss as:
\begin{flalign}
\noindent
 \mathcal{L}^{Res}_{inv}\left( G_{Res}, G_{Deg}, LR \right)=
 \mathbb{E} \left[\left\| LR^{\uparrow\downarrow} -LR \right\| _1\right] \label{res_inv_loss} 
\end{flalign}

The restoration-degradation inverse-consistency, \(G_{Res}\cdot G_{Des}: HR \rightarrow LR^{\downarrow} \rightarrow HR^{\downarrow\uparrow}\), is also deployed to facilitate more precise training as:
\begin{flalign}
\noindent
 \mathcal{L}^{Deg}_{inv}\left( G_{Deg}, G_{Res}, HR \right)=
\mathbb{E} \left[\left\| HR^{\downarrow\uparrow} -HR \right\|_1\right] 
\label{deg_inv_loss}
\end{flalign}

Furthermore, during the image translation using GAN, there is a lack of pixel-level regularization. Thus, common features shared by both $LR$ and $HR$ images will possibly be altered when the generators are over-fitted, such as the morphology of vessels. Discriminators may not be capable of distinguishing these features and consequently overlook the alternations. To alleviate it, an identity loss is introduced to \(G_{Res}\) with the input being $HR$. This identity loss is formulated as follows:
\begin{flalign}
\noindent
 \mathcal{L}^{Res}_{idt}\left(G_{Res}, HR\right)=
 \mathbb{E} \left[\left\| G_{Res}\left( HR \right)-HR \right\| _1\right]
\label{res_idt_loss}
\end{flalign}

To ensure the transitivity of the inverse workflow, we also introduce identity loss to \(G_{Deg}\) using $LR$ as:
\begin{flalign}
\noindent
 \mathcal{L}^{Deg}_{idt}\left(G_{Deg}, LR\right)=
 \mathbb{E} \left[\left\| G_{Deg}\left( LR \right) -LR \right\| _1\right]
\label{deg_idt_loss}
\end{flalign}
Identity losses preserve vessel structures and suppress unexpected noises and artifacts in generating $LR$ and $HR$ images.

\subsection{Frequency-aware Adversarial Loss}
For Generative Adversarial Network (GAN), a powerful discriminator induces the generator to produce high-quality results. As for the super-resolution, GAN aspires to produce $HR$ images by accurately enhancing $hf$ details while preserving $lf$ contours. Therefore, we facilitate the discriminator in distinguishing frequency information and thus propose the frequency-aware adversarial loss (\textbf{FAL}). 
 


\begin{figure}[t]
    \centering
    \includegraphics[scale=0.21]{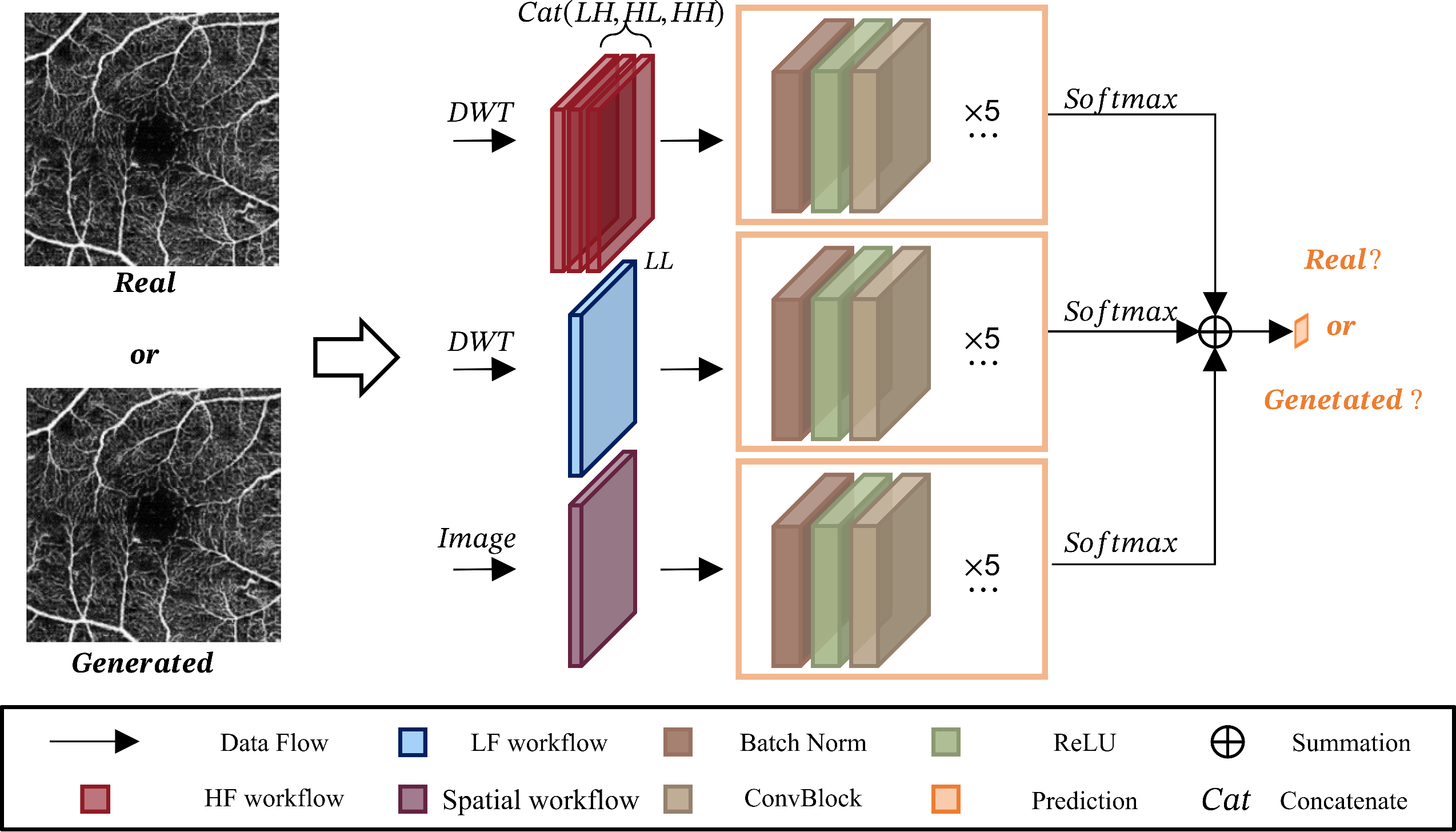}
    \caption{Structure of the discriminator. To distinguish the image as either real or generated, our method combines both frequency and spatial information in the discriminating phase. Results are aggregated to formulate the frequency-aware adversarial loss.}
    \label{fig:dis}
    \end{figure}

Inspired by the superior performance of wavelets in discriminating frequency information~\cite{wei2021unsupervised}, Haar discrete wavelets transformation (DWT) is employed. High- and low-pass filters decompose frequency components in each of the vertical and horizontal directions. Thus, as depicted in Fig. \ref{fig:dis}, decomposed components include four combinations: $LL$, $LH$, $HL$, and $HH$, where $L$ is $lf$ and $H$ is $hf$. In our framework, we refer to all ($LH$, $HL$, $HH$) as $hf$, $LL$ as $lf$, and original image as spatial information. They are simultaneously fed to three branches for feature embedding, which are finally aggregated for the final discrimination.


To optimize our restoration and degradation models in an unpaired manner, we employ two discriminators to distinguish between real and generated $LR$ and $HR$ images, denoted as \(D_{LR}\) and \(D_{HR}\), respectively. Thus, the FAL for restoration \(G_{Res}\) is defined as: 
\begin{flalign}
\label{gan_res}
\mathcal{L} _{FAL}^{Res}\left( G_{Res},D_{HR},HR \right) 
&=\mathbb{E} \left[ \left\| D_{HR}\left( HR^{\uparrow}  \right) -1 \right\| ^2 \right] \notag \\
& +\mathbb{E} \left[ \left\| D_{HR}\left( HR \right) \right\| ^2 \right]  
\end{flalign}
where \(D_{HR}\) stands for discriminator and follows the labeling scheme that real $HR$ image is 1 and restored $HR\uparrow$ image is 0. The mean square error is used, following the formulation of the least-square GAN~\cite{mao2017least}. Similarly, the FAL for the degradation \(G_{Deg}\) is defined as:
\begin{flalign}
\label{gan_deg}
\mathcal{L}_{FAL}^{Deg}\left( G_{Deg},D_{LR},LR \right) &
=\mathbb{E} \left[ \left\| D_{LR}\left( LR^{\downarrow}  \right)-1 \right\| ^2 \right] \notag \\
& +\mathbb{E} \left[ \left\| D_{LR}\left( LR \right) \right\| ^2 \right]  
\end{flalign}
According to the essential paradigm of GAN~\cite{goodfellow2020generative}, the Eq. \eqref{gan_res} and \eqref{gan_deg} are minimized to optimize generators whereas being maximized to optimize the discriminators for distinguishing between real and generated data as sensitively as possible.

\definecolor{mygray}{gray}{.9}

\begin{table*}[!t]
\centering
\setlength\tabcolsep{2pt}

\caption{Results of different methods on the CUHK-STDR of fovea-central and whole area. $\uparrow$ means the higher the better.}
\begin{tabular}{p{4.2cm}p{1.55cm}p{1.55cm}p{1.5cm}p{1.55cm}p{1.55cm}p{1.55cm}p{1.55cm}p{1.55cm}}
\toprule 
\multirow{2}{*}{Method} & \multicolumn{4}{l}{fovea-central area} & \multicolumn{4}{l}{whole area}   \\ 
\cmidrule(l){2-5}\cmidrule(l){6-9}
                    & PSNR$\uparrow$ & SSIM$\uparrow$ & NMI$\uparrow$ & FSIM$\uparrow$ & PSNR$\uparrow$ & SSIM$\uparrow$ & NMI$\uparrow$ & FSIM$\uparrow$ \\ 
\hline
CycleGAN-ResNet~\cite{zhu2017unpaired,he2016deep} & $16.827_{0.162}$ & $0.481_{0.003}$ & $1.054_{0.001}$ & $0.639_{0.003}$ & $17.183_{0.108}$ & $\underline{0.544_{0.002}}$ & $1.059_{0.001}$ & $0.663_{0.001}$ \\

CinCGAN~\cite{yuan2018unsupervised} & $16.991_{0.274}$ & $0.502_{0.025}$ & $1.052_{0.001}$ & $\underline{0.677_{0.014}}$ & $17.177_{0.231}$ & $0.532_{0.013}$ & $1.056_{0.002}$ & $0.673_{0.005 }$ \\

Pseudo-Supervision~\cite{maeda2020unpaired} & $17.193_{0.109}$ & $0.446_{0.008}$ & $1.054_{0.000}$ & $0.655_{0.002}$ & $17.611_{0.016}$ & $0.514_{0.009}$ & $1.062_{0.000}$ & $0.678_{0.002}$ \\

DA-Unsupervised SR~\cite{wang2021unsupervisedDA} & $16.426_{0.073}$ & $0.426_{0.002}$ & $1.055_{0.000}$ & $0.616_{0.000}$ & $17.523_{0.197}$ & $0.513_{0.002}$ & $\mathbf{1.064_{0.001}}$ & $0.651_{0.003}$ \\

Frequency GAN~\cite{zhang2022frequency} & $\underline{17.435_{0.046}}$ & $\underline{0.507_{0.003}}$ & $\underline{1.057_{0.001}}$ & $0.667_{0.004}$ & $\underline{17.547_{0.054}}$ & $0.529_{0.004}$ & $1.061_{0.001}$ & $\underline{0.679_{0.004}}$ \\

\rowcolor{mygray}\textbf{FUSR} & $\mathbf{17.487_{0.034}}$ & $\mathbf{0.512_{0.007}}$ & $\mathbf{1.057_{0.001}}$ & $\mathbf{0.679_{0.004}}$ & $\mathbf{17.679_{0.072}}$ & $\mathbf{0.555_{0.002}}$ & $\underline{1.063_{0.000}}$ & $\mathbf{0.689_{0.003}}$ \\
\bottomrule
\end{tabular}
\label{table:perform}

\end{table*}


\definecolor{mygray}{gray}{.9}

\begin{table*}[!ht]
\centering
\setlength\tabcolsep{2pt}

\caption{Results of the experiments on frequency components on the CUHK-STDR of fovea-central and whole area. $\uparrow$ means the higher the better, while $\downarrow$ means the lower the better.}
\begin{tabular}{p{1.0cm}p{1.55cm}p{1.55cm}p{1.55cm}p{1.55cm}p{1.55cm}p{1.55cm}p{1.55cm}p{1.55cm}p{1.55cm}p{1.55cm}}
\toprule 
\multirow{2}{*}{Setting} & \multicolumn{5}{l}{fovea-central area} & \multicolumn{5}{l}{whole area}   \\ 
\cmidrule(l){2-6}\cmidrule(l){7-11}
& PSNR$\uparrow$ & SSIM$\uparrow$ & NMI$\uparrow$ & FSIM$\uparrow$ & LFD$\downarrow$ & PSNR$\uparrow$ & SSIM$\uparrow$ & NMI$\uparrow$ & FSIM$\uparrow$  & LFD$\downarrow$\\
\hline
$hf$ only & 
$\underline{17.230_{0.081}}$ & $\mathbf{0.515_{0.007}}$ & $\underline{1.057_{0.001}}$ & $\underline{0.675_{0.007}}$ & $18.240_{0.068}$ & $17.403_{0.096}$ & $\mathbf{0.556_{0.002}}$ & $\underline{1.062_{0.001}}$ & $\underline{0.688_{0.003}}$ & $18.258_{0.095}$ \\

$lf$ only & $16.924_{0.076}$ & $0.393_{0.014}$ & $1.053_{0.001}$ & $0.584_{0.027}$  &$\underline{18.221_{0.024}}$ & $\underline{17.461_{0.082}}$ & $0.473_{0.018}$  & $1.062_{0.001}$ & $0.631_{0.013}$ & $\underline{18.156_{0.005}}$ \\

\rowcolor{mygray}\textbf{FUSR} & $\mathbf{17.487_{0.034}}$ & $\underline{0.512_{0.007}}$ & $\mathbf{1.057_{0.001}}$ & $\mathbf{0.679_{0.004}}$ & $\mathbf{18.150_{0.014}}$ & $\mathbf{17.679_{0.072}}$ & $\underline{0.555_{0.002}}$ & $\mathbf{1.063_{0.000}}$ & $\mathbf{0.689_{0.003}}$ & $\mathbf{18.124_{0.007}}$ \\
\bottomrule
\end{tabular}
\label{table:frequencies}

\end{table*}

\definecolor{mygray}{gray}{.9}
\begin{table*}[!h]
    \centering
    \setlength\tabcolsep{2pt}

    \caption{Results of the ablation studies of loss functions on the CUHK-STDR of fovea-central and whole area. $\uparrow$ means the higher the better. $\checkmark$ means  "included".}
    \begin{tabular}{p{0.7cm}p{1cm}p{1cm}p{0.7cm}p{1cm}p{1.5cm}p{1.5cm}p{1.5cm}p{1.5cm}p{1.5cm}p{1.5cm}p{1.5cm}p{1.5cm}}
    \toprule 
    
    \multicolumn{3}{l}{Consistency loss} & \multicolumn{2}{l}{Adversarial loss} & \multicolumn{4}{l}{fovea-central area} & \multicolumn{4}{l}{whole area}   \\ 
    
    \cmidrule(l){1-3} \cmidrule(l){4-5}\cmidrule(l){6-9}\cmidrule(l){10-13}
    \textbf{FFCL} & ordinary & FFL~\cite{jiang2021focal} & \textbf{FAL} & ordinary & PSNR$\uparrow$ & SSIM$\uparrow$ & NMI$\uparrow$ & FSIM$\uparrow$ & PSNR$\uparrow$ & SSIM$\uparrow$ & NMI$\uparrow$ & FSIM$\uparrow$  \\
    \cmidrule(l){1-3} \cmidrule(l){4-5}\cmidrule(l){6-9}\cmidrule(l){10-13}
    \rowcolor{mygray}\checkmark&&&\checkmark&& 
    $\mathbf{17.487_{0.034}}$ & $0.512_{0.007}$ & $\mathbf{1.057_{0.001}}$ & $\underline{0.679_{0.004}}$ & $\mathbf{17.679_{0.072}}$ & $\underline{0.555_{0.002}}$ & $\mathbf{1.063_{0.000}}$ & $\mathbf{0.689_{0.003}}$ \\

    \checkmark&&&&\checkmark& 
    $17.335_{0.044}$ & $\mathbf{0.516_{0.003}}$ & $1.055_{0.000}$ & $0.671_{0.001}$ & $17.393_{0.019}$ & $0.552_{0.002}$ & $1.060_{0.000}$ & $0.685_{0.002}$ \\
    
    &\checkmark&&\checkmark&& 
    $17.172_{0.045}$ & $0.409_{0.016}$ & $1.055_{0.001}$ & $0.603_{0.013}$ & $\underline{17.610_{0.076}}$ & $0.480_{0.006}$ & $\underline{1.063_{0.000}}$ & $0.638_{0.003}$ \\

    &\checkmark&&&\checkmark& 
    $16.827_{0.162}$ & $0.481_{0.003}$ & $1.054_{0.001}$ & $0.639_{0.003}$ & $17.183_{0.108}$ & $0.544_{0.002}$ & $1.059_{0.001}$ & $0.663_{0.001}$ \\

    &&\checkmark&\checkmark&& 
    $\underline{17.442_{0.030}}$ & $\underline{0.515_{0.001}}$ & $\underline{1.056_{0.000}}$ & $\mathbf{0.680_{0.002}}$ & $17.573_{0.090}$ & $\mathbf{0.557_{0.003}}$ & $1.062_{0.001}$ & $\underline{0.689_{0.001}}$  \\

    \bottomrule
    \end{tabular}
    \label{table:ablation}

    \end{table*}

\subsection{Frequency-aware Focal Consistency Loss}
To preserve the spectrum distribution in the reconstructed results of $G_{Res}$ and $G_{Deg}$, we introduce a frequency-aware focal consistency loss (\textbf{FFCL}). Given the limitations of neural networks and GANs in accurately capturing $hf$, this term aims to enforce consistency in frequency information and place additional emphasis on $hf$.

Specifically, a spectrum weighting matrix~\cite{jiang2021focal} is first formulated to penalize the spectral consistency error as:
\begin{flalign}
 w\left( u,v \right) =\left| \mathcal{X}'\left( u,v \right) - \mathcal{X}\left( u,v \right) \right|^{\gamma_1}
\label{focalweight}
\end{flalign}
where $\mathcal{X}$ and $\mathcal{X}'$ are the frequency representation of the original and reconstructed images $x$ and $x'$ via Eq. \eqref{fft}. $\gamma_1$ is the scaling factor. Then, the loss to ensure the consistency of the restoration-degradation as:
\begin{flalign}
\label{FFCL}
\mathcal{L}_{FFCL}(x,x')
&=\frac{1}{MN}\sum_{u=0}^{M-1}\sum_{v=0}^{N-1} w_{hf}\odot \left| \mathcal{X}'_{hf} -\mathcal{X}_{hf} \right|^2 \notag \\
& + \gamma_2w_{lf}\odot \left| \mathcal{X}'_{lf} -\mathcal{X}_{lf} \right|^2 
\end{flalign}
where $u$ and $v$ are coordinates of $\mathcal{X}$ anf $\mathcal{X}'$. $w$ represents the spectrum weighting matrix using Eq. \eqref{focalweight}. The subscripts $hf$ and $lf$ indicate the $hf$ and $lf$ components, respectively. $\gamma_2$ is a scaling factor. The symbol $\odot$ denotes the Hadamard product, which applies the weighting matrix $w$ to the mean square error between the spectra $\mathcal{X}$ and $\mathcal{X}'$.

By aggregating all terms of our proposed loss functions, we form the final objective $\mathcal{L}_{Total}$ as:
\begin{flalign}
\label{genloss}
\min_{G}\max_{D}\mathcal{L}_{Total}
&=\left(\mathcal{L}^{Deg}_{FAL}+\beta_1\mathcal{L}^{Res}_{FAL}\right) +\left(\mathcal{L}^{Deg}_{inv}+\beta_2\mathcal{L}^{Res}_{inv}\right) \notag \\
& +\left(\mathcal{L}^{Deg}_{idt}+\beta _3\mathcal{L}^{Res}_{idt}\right) + \beta _4\mathcal{L}_{FFCL} 
\end{flalign}
where $G$ comprises both $G_{Res}$ and $G_{Deg}$, while $D$ consists of $D_{HR}$ and $D_{LR}$. Then, the total loss $\mathcal{L}_{Total}$ encompasses adversarial loss $\mathcal{L}_{\textbf{FAL}}$, the consistency loss ${L}_{inv}$, the identity loss $\mathcal{L}_{idt}$, and the frequency-aware focal consistency loss $\mathcal{L}_{\textbf{FFCL}}$. The parameters $\beta_1$, $\beta_2$, $\beta_3$, and $\beta_4$ in Eq. \eqref{genloss} are empirically set to 1, 10, 5, and 1, respectively. In the training phase, Eq. \eqref{genloss} is minimized to optimize the generators, while it is maximized to optimize the discriminator iteratively.

In summary, we propose a GAN-based unpaired OCTA super-resolution. By leveraging spatial and frequency information, we improve the resolution while preserving $hf$. Our dual-path generators separately refine $hf$ while retaining $hf$ components, and the discriminators incorporate spatial and wavelets information as the FAL. We also introduce the FFCL to dynamically preserve the entire spectral consistency.

\begin{figure*}[!h]
    \centering
    \includegraphics[scale=0.2]{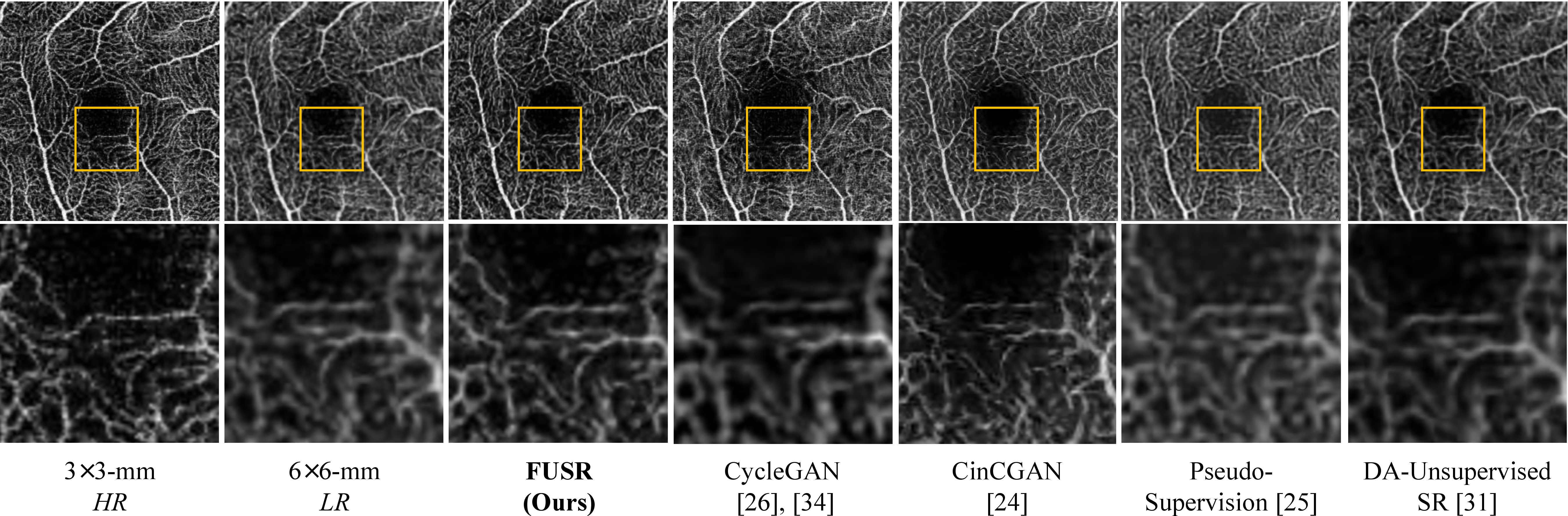}
    \caption{Visual results in fovea-central area. The first row illustrated exemplary original and reconstructed results. The second row illustrates zoomed details.}
    \label{fig:results}
\end{figure*}
\definecolor{mygray}{gray}{.9}

\begin{table*}[!ht]
\centering
\setlength\tabcolsep{2pt}

\caption{Results of the experiments on frequency-aware discriminators on the CUHK-STDR of fovea-central and whole area. $\uparrow$ means the higher the better, while $\downarrow$ means the lower the better.}
\begin{tabular}{p{4.2cm}p{1.55cm}p{1.55cm}p{1.5cm}p{1.55cm}p{1.55cm}p{1.55cm}p{1.55cm}p{1.55cm}}
\toprule 
\multirow{2}{*}{Method} & \multicolumn{4}{l}{fovea-central area} & \multicolumn{4}{l}{whole area}   \\ 
\cmidrule(l){2-5}\cmidrule(l){6-9}
                    & PSNR$\uparrow$ & SSIM$\uparrow$ & NMI$\uparrow$ & FSIM$\uparrow$ & PSNR$\uparrow$ & SSIM$\uparrow$ & NMI$\uparrow$ & FSIM$\uparrow$ \\ 
\hline

Ordinary & $17.172_{0.045}$ & $0.409_{0.016}$ & $1.055_{0.001}$ & $0.603_{0.013}$ & $17.610_{0.076}$ & $0.480_{0.006}$ & $\underline{1.063_{0.000}}$ & $0.638_{0.003}$ \\

Fourier & 
$\underline{17.381_{0.055}}$ & $\mathbf{0.517_{0.0229}}$ & $\underline{1.057_{0.001}}$ & $\underline{0.674_{0.006}}$ & $\underline{17.658_{0.138}}$ & $\mathbf{0.559_{0.0085}}$ & 
$ 1.063_{0.001}$ & $\underline{0.684_{0.006}}$ \\

\rowcolor{mygray}\textbf{FUSR (Wavelets)} & $\mathbf{17.487_{0.034}}$ & $\underline{0.512_{0.007}}$ & $\mathbf{1.057_{0.001}}$ & $\mathbf{0.679_{0.004}}$ & $\mathbf{17.679_{0.072}}$ & $\underline{0.555_{0.002}}$ & $\mathbf{1.063_{0.000}}$ & $\mathbf{0.689_{0.003}}$ \\
\bottomrule
\end{tabular}
\label{table:wavelets}

\end{table*}

\section{Experimental Evaluation}
\subsection{Dataset}
The OCTA images used in this work were retrospectively collected from the Chinese University of Hong Kong Sight-THReatening Diabetic Retinopathy (CUHK-STDR) study. This study was an observational clinical study focused on diabetic retinal disease in subjects with Type 1 or Type 2 Diabetes Mellitus recruited from the CUHK Eye Centre and Hong Kong Eye Hospital~\cite{Yangbjophthalmol-2021-320779, sun2019oct, tang2017determinants,yang2023assessment}. The OCTA imaging was performed using a swept-source optical coherence tomography (DRI OCT Triton; Topcon, Tokyo, Japan), which is one of the mainstream OCT devices~\cite {lu2021quantitative}. Notably, we leveraged paired images in CUHK-STDR dataset to train the models in an unpaired manner while evaluating performances using pairwise metrics.

Specifically, 296 pairs of fovea-central $HR$ and $LR$ OCTA images (see Fig. \ref{data}) were collected and split in the proportion of 4:1 for training and validation. Otherwise, an additional 279 groups of paired $HR$ and $LR$ images for the whole area were also purposely collected and used for testing. Specifically, each group consisted of one $LR$ and five $HR$. $HR$ images included one fovea-center and four parafovea, which were combined to generate a whole $HR$ 6mm$\times$6mm montage registered to the original $LR$. It is also worth noting that most typical OCTA images are fovea-centered since it is the region of interest in various retinal disease studies~\cite{hwang2015optical,
roisman2016optical,sun2019oct,che2022learning,che2023dgdr,yang2023assessment}. Thus, we only used fovea-centered images for training while including parafoveal images for testing to align with real-world clinics.

In the preprocessing of the training set, the original $LR$ images were upsampled using bicubic interpolation (Fig. \ref{data}. A). We then cropped 256$\times$256 patches from the upsampled $LR$ images (Fig. \ref{data}. B) and corresponding $HR$ images (Fig. \ref{data}. C). During the training phase, these cropped $LR$ and $HR$ patches were randomly selected and provided to the network in an unpaired manner. To prepare a pixel-wise aligned testing dataset for quantitative evaluation, each $LR$ image was paired with an $HR$ image from the same eye of the same patient. To account for slight structural changes due to the time interval between capturing the images, registration was performed to align the paired images. Thus, to evaluate the performance, the paired images were provided to the model after proper image registration.

\begin{figure*}[!t]
\centering
\includegraphics[scale=0.229]{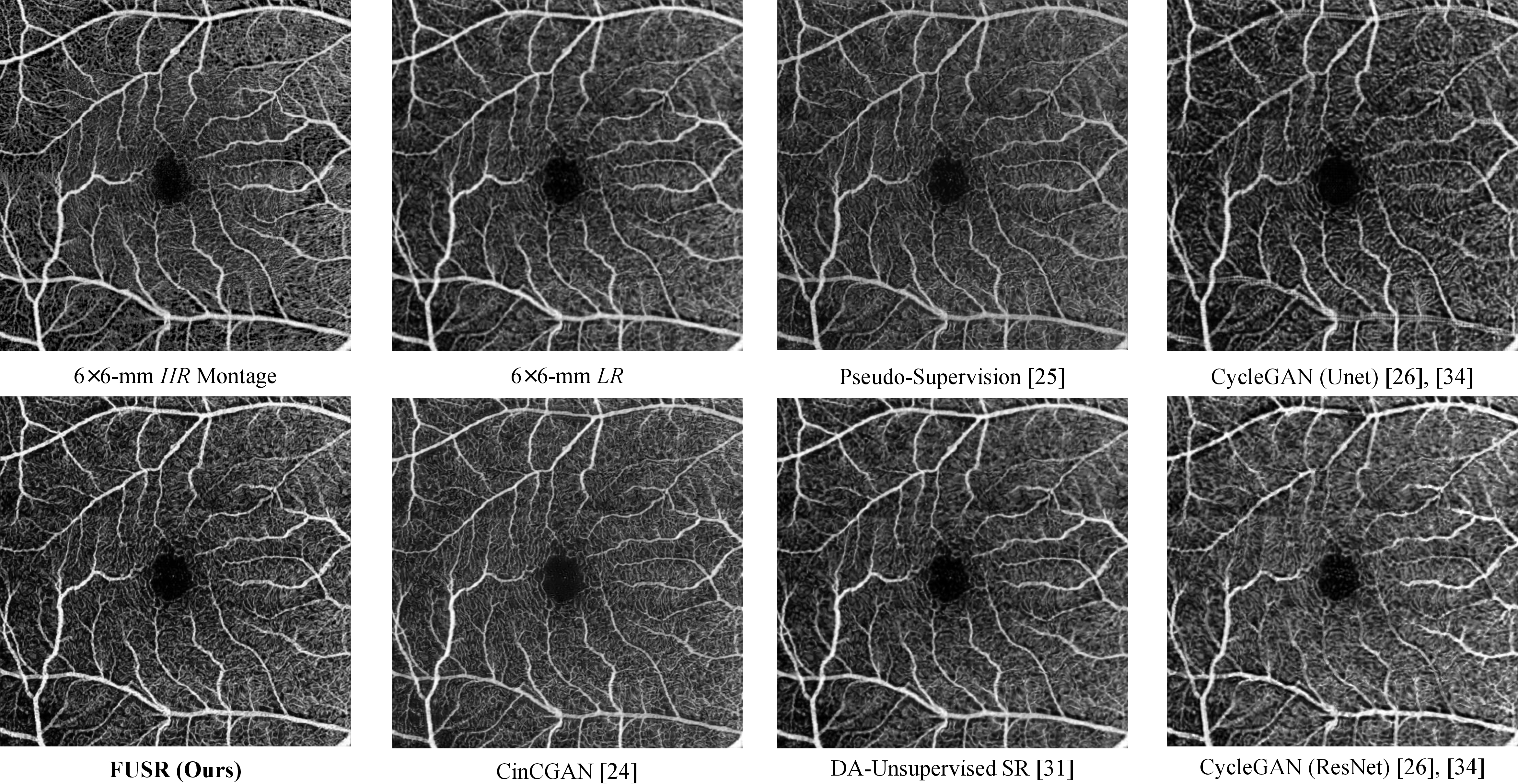}
\caption{ Visual results of the exemplary original and reconstructed images on the whole 6mm$\times$6mm OCTA image.}
\label{fig:results_whole}
\end{figure*}

\subsection{Implementation Details}
Our model was trained on one NVIDIA RTX 3090 with 24GB memory. The parameters were initialized using the standard normal distribution. The initial learning rate was set to 0.0002, and it decayed linearly to 0 during training. The training phase optimized the parameters for a minimum of 5,000 iterations. In each iteration, an unaligned pair of $HR$ and $LR$ images was provided for the network. Specifically, the $HR$ image was a 3mm$\times$3mm patch cropped with 256-pixel$\times$256-pixel, while the $LR$ image was a 6mm$\times$6mm patch upsampled using bicubic interpolation and then randomly cropped with 256-pixel$\times$256-pixel.

We evaluated the performance using common pixel-wise paired metrics in super-resolution studies. Specifically, we used peak signal-to-noise ratio (PSNR), structural similarity index measure (SSIM)~\cite{wang2004image}, normalized mutual information (NMI), and feature similarity index measure (FSIM)~\cite{zhang2011fsim}. PSNR measured valid signals compared to noises. SSIM evaluated quality in terms of structure, illuminance, and contrast. NMI evaluated how matched the two images were. FSIM was a feature-level frequency-aware measurement that considered phase congruency and gradient magnitude.

\begin{figure*}[!t]
    \centering
    \includegraphics[scale=.18]{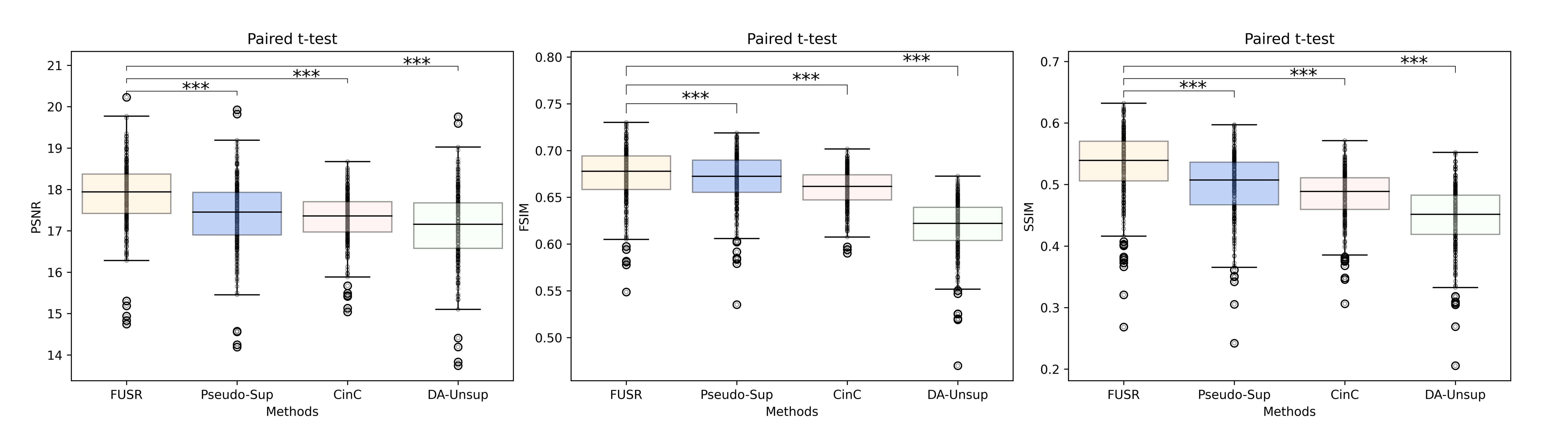} 
    \caption{Boxplots of quantitative performance comparisons across multiple methods over the fovea-central dataset. The asterisks denote the \textit{p}-value in the paired t-test, where '-' stands for no statistical significance, $\ast$ for \textit{p} $<$ 0.05, $\ast\ast$ for \textit{p} $<$ 0.01, $\ast\ast\ast$ for \textit{p} $<$ 0.001.}
    \label{figure:ttest}
\end{figure*} 
\subsection{Experimental Results and Comparison}
In Table. \ref{table:perform}, we compare our method to several baselines, including CycleGAN~\cite{zhu2017unpaired}, Cycle-in-Cycle GAN (CinCGAN)~\cite{yuan2018unsupervised}, Pseudo-Supervision~\cite{maeda2020unpaired}, and Domain Adaptation Unsupervised Super-Resolution (DA-Unsupervised SR)~\cite{wang2021unsupervisedDA}. We also reimplement our previous work~\cite{zhang2022frequency}. Similar to our approach, these baseline methods are all in the unpaired setting. The quantitative results demonstrate that our method outperforms them and also surpasses our previous work. Moreover, although $HR$ peripheries are unseen during training, our experiments on the whole area still show improved resolution.

Fig. \ref{fig:results} presents the visual results of our method. It showcases how our approach improves the resolution of OCTA images while preserving fine capillary structures. In comparison, CinCGAN introduces noise and loses the original vessel features. Pseudo-Supervision and DA-Unsupervised SR exhibit lower contrast compared to our method. CycleGAN successfully recovers $hf$ vasculature but it disrupts the vessel coherence of the foveal avascular zone (FAZ). On the other hand, our method refines $hf$ information while introducing minimal unexpected noise and retaining most of the original information. Notably, CinCGAN visually resembles the $HR$ ground truth images (Fig. \ref{fig:results_whole}) with high SSIM and FSIM. However, the cost is the signal-to-noise ratio, as shown in Table. \ref{table:perform}. Therefore, our proposed approach achieves a better balance between structural information and the signal-to-noise ratio, resulting in overall higher fidelity.

Furthermore, as depicted in Fig. \ref{figure:ttest}, we conducted the paired t-tests over our method with other state-of-the-art methods on the whole fovea-central dataset. The \textit{p}-values of each t-test are less than 0.001, statistically indicating that our method significantly outperforms the baseline methods in the metrics of PSNR, SSIM, and FSIM. Results further imply that our method could better refine the recovering structural information without introducing strong noises.

\subsection{Frequency Decomposition in Generators}
As previously illustrated in Fig. \ref{fig:pipeline}, we decompose $hf$ and $lf$ in generators to exceptionally enhance fine-grained details and alleviate $lf$-bias. Specifically, $hf$ is provided for neural networks via $HFB$ operation. Thus we further investigate the effectiveness of frequency decomposition.

First, as shown in Table. \ref{table:frequencies}, we respectively remove $hf$ and $lf$ by replacing frequency components with tensors filled with all zeros and freezing the gradients. Results show that either removing $hf$ or $lf$ would degrade the performance with respect to all metrics. Comparably, removing $hf$ would lead to more severe degradation, especially in terms of FSIM. This confirms the $lf$-bias of the neural networks and verifies the effectiveness of our method. 

Aside from evaluation in the spatial domain, we also examine the frequency domain by employing the Log Frequency Distance (LFD) metric \cite{jiang2021focal}, which is formed as:
\begin{flalign}
    LFD = \mathrm{log} \left[ \frac{1}{MN}\left( \sum_{m=0}^{M-1}{\sum_{n=0}^{N-1}{\left| F_x -F_y \right|^2}} \right)+1 \right]
   \label{LFD}
\end{flalign}
where \(F_x\) and \(F_y\) are spectrums of generated and real images via Eq. \eqref{fft}. $m$ and $n$ are the Cartesian coordinates of the $F$. Eq. \eqref{LFD} measures the log mean square error over the whole spectrums. As shown in Table. \ref{table:frequencies}, the reconstruction from components only with $hf$ would lead to higher spectral errors than results from $hf$ only. This may be attributed to the common view that most spectral powers are concentrated in $lf$ bands~\cite{chandrasegaran2021closer}. Thus, although the spectral errors for $hf$ are lower than that for $lf$, these slight disturbances for $hf$ would lead to more severe quality degradation in spatial domain images. Results verify the necessity to exceptionally emphasize $hf$ while preserving $lf$.

\begin{figure}[!t]
    \includegraphics[scale=.30]{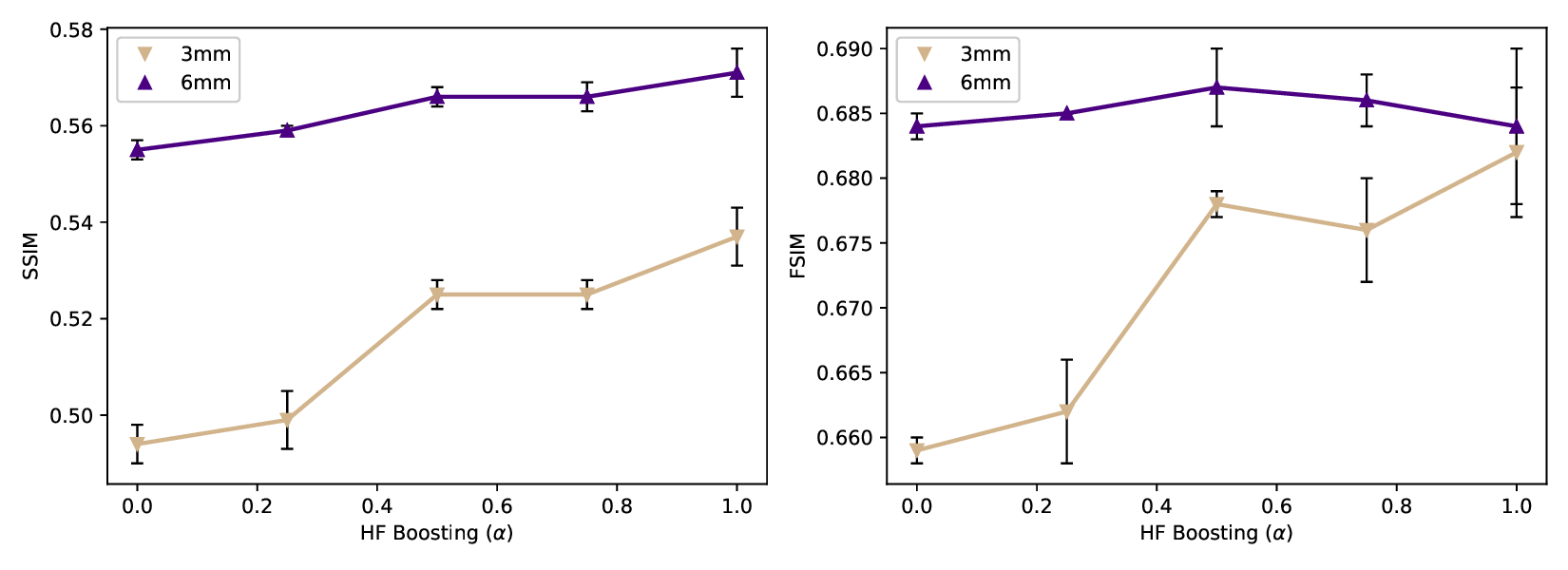} 
    \caption{Plots of the $HFB$ rates versus different qualitative metrics.}
    \label{figure:hfb}
\end{figure} 
Moreover, we also evaluate the effectiveness of $HFB$ in Eq. \eqref{hfb}. As shown in Fig. \ref{figure:hfb}, results illustrate the relationship between qualitative performance and the level of boosting, which is $\alpha$ in Eq. \eqref{hfb}. We observe better performances in terms of SSIM and FSIM as more $hf$ information is provided. This indicates that $hf$ components play a crucial role in preserving structural information. The above results demonstrate the significance of our $HFB$ operation in frequency decomposition.

\begin{figure}[!t]
\centering
    \includegraphics[scale=.205]{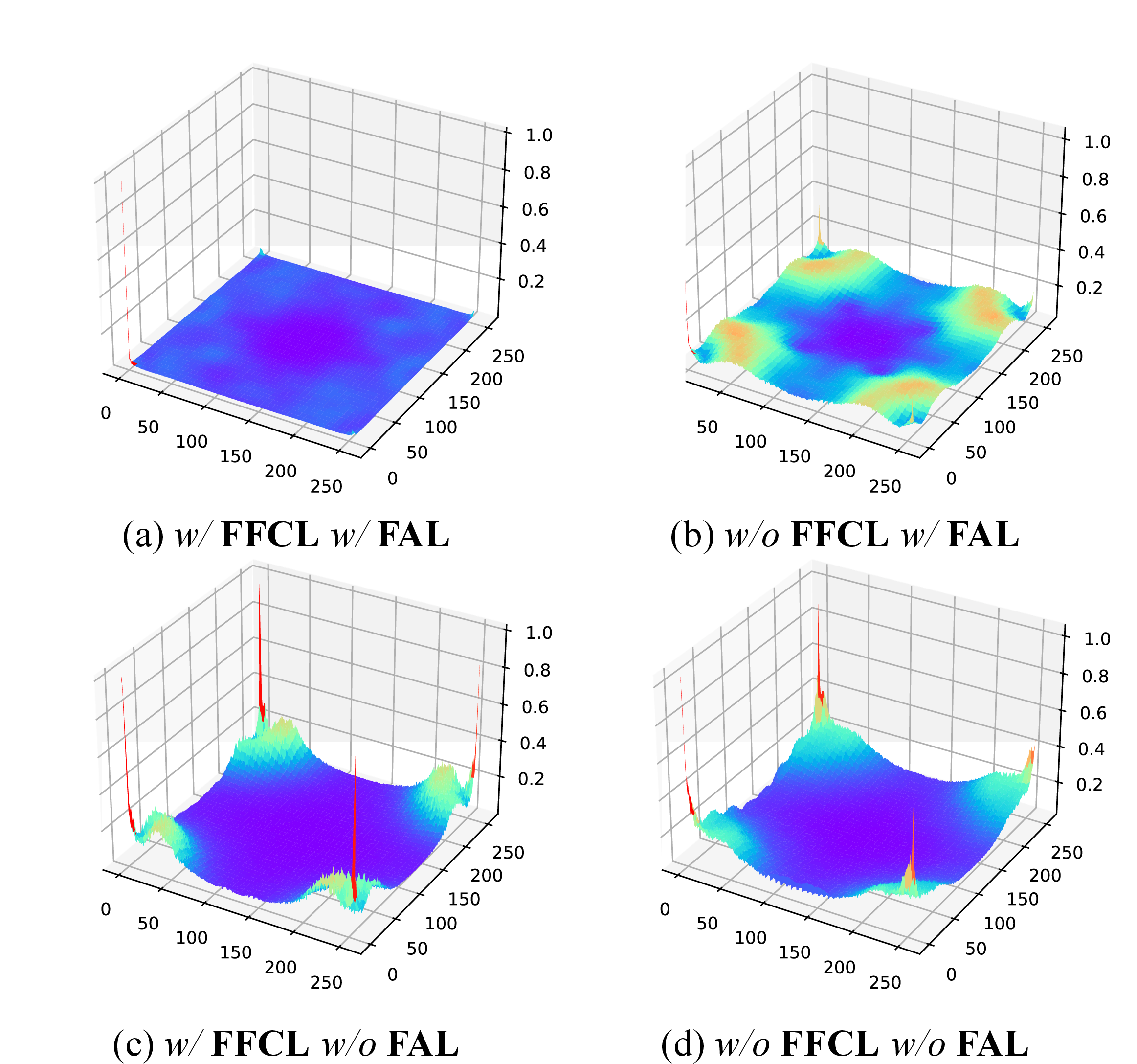} 
    \caption{3D visualization for spectral errors. The horizontal plane is the spectrum coordinates, while the vertical axis indicates the error intensity. Larger errors indicate less accurately reconstructed frequencies.}
    \label{figure:Spec_error}
\end{figure} 

\subsection{Frequency-Aware Loss Functions}
We introduce two frequency-aware loss functions: frequency-aware focal consistency loss (\textbf{FFCL}) and frequency-aware adversarial loss (\textbf{FAL}). They were designed to provide more precise supervision over the spectrum. To verify the effectiveness, we present the 3D visualization for errors in the spectrum for different results obtained using (FFCL) and FAL, as shown in Fig. \ref{figure:Spec_error}. It can be observed that without FFCL and FAL, most of the spectral errors are introduced by the $hf$ components. This indicates the importance of addressing the frequency information for accurate spectral reconstruction. FAL enables the discriminators to be more aware of the $hf$ components and reduces the corresponding reconstruction errors. However, there are still inaccuracies in the middle- to high-frequency components. Otherwise, FFCL retains more precise middle-frequency information but lacks the capability to reproduce $hf$ accurately. Consequently, by leveraging both FFCL and FAL, our super-resolution model is able to maintain precise results across the entire spectrum. This approach effectively addresses the challenges posed by different frequency components and enhances spectral precision.

Moreover, we evaluate the efficiency of our DWT-based FAL. Previous work~\cite{wei2021unsupervised} indicated superior performance in discriminating frequency information. We further compared it with the FFT-based and ordinary discriminators in Table. \ref{table:wavelets}. Results indicate that leveraging frequency information could improve the super-resolution quality, while the DWT-based design shows better capability to improve the image quality than the FFT-based method.

\subsection{Ablation Study}
To evaluate the effects of different components, we continue to conduct ablation on FAL and FFCL, as shown in Table. \ref{table:ablation}. The row highlighted in gray is our method with FAL and FFCL. By replacing our FAL and FFCL with ordinary adversarial loss and consistency loss functions, we examined the contribution of our designs. Compared with ordinary adversarial loss, our FAL better compresses the level of noise. Meanwhile, compared with ordinary consistency loss, FFCL improves the quality in all metrics. Generally, we observed a decline or similar performance in PSNR, particularly in the fovea-central area. This indicates that the frequency-aware losses applied to the generators and discriminators play a crucial role in controlling the noise intensity. These findings are consistent with the visualizaion in Fig. \ref{fig:results}.

Furthermore, we replaced the proposed FFCL with the ordinary focal frequency loss (\textbf{FFL})~\cite{jiang2021focal} to evaluate our design in the unpaired super-resolution setting. Comparing the results, we found that FFL can preserve the structural information related to $hf$ components, but our proposed FFCL achieves higher accuracy in preserving the structural details. Meanwhile, FFCL also enables the model to better generalize to the whole 6mm$\times$6mm area, where lack of the peripheral regions as the supervision signal. These results empirically evidence the effectiveness of each component in our approach, showcasing contributions to reconstructing high-quality images in both the spatial and frequency domains.

\section{Conclusion}

This paper introduced Frequency-aware Unpaired Super-Resolution (FUSR) for OCTA images, enhancing resolution in an unpaired setting. We proposed a GAN-based framework to mimic restoration and degradation mappings, optimized end-to-end through consistency loss. Recognizing the importance of fine-grained capillaries in microvasculature as biomarkers for OCTA, we employed frequency decomposition to emphasize high-frequency (hf) components by separating and fusing frequency elements. We also introduced a Frequency-Aware Loss (\textbf{FAL}) for the discriminators to better preserve capillary structure and a Frequency-Feature Consistency Loss (\textbf{FFCL}) to maintain spectrum consistency.

Our experiments and analytical studies validated the method’s effectiveness and demonstrated superior performance. To the best of our knowledge, as an extension of our previous work~\cite{zhang2022frequency}, our studies were the first to leverage frequency analysis and utilize GANs in unpaired OCTA super-resolution. This approach addressed challenges associated with large-scale data collection and complex data preparation required in conventional supervised super-resolution methods from a frequency-domain analysis perspective.

However, the unpaired super-resolution setting lacks pixel-wise supervision, potentially resulting in performance that may not match fully supervised approaches~\cite{article1,article2,hao2022sparse}. As shown in Fig. 6, while our method enhances hf capillary details without introducing extra artifacts, it cannot precisely infer missing semantic information, possibly leading to vessel incoherence. Another limitation of current approaches, including ours, is the assumption that features inside and outside the fovea-central 3mm×3mm area are identically distributed, as only this region provides high-resolution supervision information. This assumption may impede the model’s generalizability. A potential solution is to develop a self-supervised learning approach for more robust learning~\cite{pathak2016context}.

Furthermore, clinical validation will be conducted across multiple OCT devices in our future work~\cite{yang2023deep}. Currently, there is a lack of other available paired OCTA image datasets enabling unpaired super-resolution study. The various operational parameters of the hardware may lead to different imaging quality and consequently affect performance and generalizability. Thus, collecting large-scale multi-center datasets for a more comprehensive study with more downstream tasks will be our future pathway.

\section*{References}
\bibliographystyle{IEEEtran}
\bibliography{refs}

\end{document}